\documentclass[12pt]{article}
\usepackage[text={6.5in,8in}, top=1.75in, left=0.95in]{geometry}

\usepackage{natbib}
\usepackage{algorithm}
\usepackage{algpseudocode}
\usepackage{booktabs}
\usepackage[figuresright]{rotating}
\usepackage{subcaption}
\usepackage{authblk}

\usepackage{amsfonts}
\usepackage{amsmath}
\usepackage{color}
\newcommand{\ba}{\begin{eqnarray}}
\newcommand{\ea}{\end{eqnarray}}
\newcommand{\bas}{\begin{eqnarray*}}
\newcommand{\eas}{\end{eqnarray*}}
\newcommand{\ben}{\begin{enumerate}}
\newcommand{\een}{\end{enumerate}}

\title{Conformal predictive  intervals in survival analysis: a re-sampling approach}

\author[1,*]{Jing Qin}
\author[2]{Jin Piao}
\author[3]{Jing Ning}
\author[3,*]{Yu Shen}

\affil[1]{\small Biostatistics Research Branch, National Institute of Allergy and Infectious Diseases, MD, U.S.A.}
\affil[2]{Department of Population and Public Health Sciences, University of Southern California, CA, U.S.A.}
\affil[3]{Department of Biostatistics, The University of Texas M.D. Anderson Cancer Center, TX, U.S.A.}
\affil[*]{Corresponding authors: \texttt{jingqin@niaid.nih.gov, yshen@mdanderson.org}}

\date{}  

\begin{document}

\maketitle

\begin{abstract}
The distribution-free method of conformal prediction \citep{vovk2005algorithmic} has gained considerable attention in computer science, machine learning, and statistics.
\cite{candes2023conformalized} extended this method to right-censored survival data, addressing right-censoring complexity by creating a covariate shift setting, extracting a subcohort of subjects with censoring times exceeding a fixed threshold. Their approach only estimates the lower prediction bound for type I censoring, where all subjects have available censoring times regardless of their failure status.
In medical applications, we often encounter more general right-censored data, observing only the minimum of failure time and censoring time. Subjects with observed failure times have unavailable censoring times. To address this, we propose a bootstrap method to construct one- as well as two-sided conformal predictive   intervals for general right-censored survival data under different working regression models.
Through simulations, our method demonstrates excellent average coverage for the lower    bound and good coverage for the two-sided  predictive interval, regardless of working model is correctly specified or not, particularly under moderate censoring. We further extend the proposed method to several  directions in medical applications. We apply this method to predict breast cancer patients' future survival times based on tumour characteristics and treatment.
\end{abstract}


\section{Introduction}
Conformal prediction has garnered significant attention and continues to grow in importance for estimating prediction intervals without relying on correctly specified model assumptions between outcomes and covariates \citep{angelopoulos2021gentle}.
In cancer research, there is a strong interest in providing predictive  intervals for survival outcomes based on patients' baseline risk factors and treatments received.
Inflammatory breast cancer (IBC) is a rare subtype, accounting for less than 5\% of all breast cancer cases. It is characterised by aggressive cancer cells that block lymph vessels in the breast skin. IBC often progresses rapidly, with all cases being diagnosed as stage III or IV disease, carrying a poor prognosis. Age, race, tumour characteristics, and treatments received after diagnosis are important risk factors associated with the survival of IBC patients. 
Predicting the prognosis or survival time of an IBC patient is crucial for both the patient and her physicians in selecting the most appropriate initial course of treatment and managing the disease optimally. While it is widely recognised that tumour characteristics, such as nodal status and stage at diagnosis, play a significant prognostic role in predicting long-term outcomes for non-IBC breast cancer patients, the impact of these tumour characteristics and demographic factors on the prognosis and response to established therapies in IBC patients has not been well-studied. Existing literature suggests that all subtypes of IBC have poor overall survival and a worse prognosis compared to their non-IBC counterparts of the same subtypes \citep{masuda2014long}. Accurately predicting overall survival time (OS) based on tumour characteristics and treatment information using the largest IBC dataset, the National Cancer Database (NCDB), could generate hypotheses for future prospective trials aiming to improve the overall survival of IBC patients.

Building on recent conformal inference literature, we propose a distribution-free predictive   interval for right-censored data in regression settings. One major advantage for the conformal inference is that it does not require a correct model assumption for either the joint or marginal distribution of $(T,{\pmb X})$ where $T$ represents the survival outcome and ${\pmb X}$ is a set of covariates. In contrast, conventional predictive models rely on a correct model assumption of the outcome given the covariates $(T|{\pmb X})$.  The essence of ``model assumption free” in conformal inference is that the conformal prediction interval always has a valid average coverage probability under a working regression model, which does not have to be correctly specified. In conformal inference, the average coverage probability refers to the probability that the true value of the survival time falls within the predictive interval. A valid predictive interval is one that has the correct average coverage probability over repeated samples. 
Note that the predictive interval is constructed based on a working model that may not accurately reflect the true relationship between the survival time and the covariates. The width of the predictive interval depends on how well the working model approximates the true relationship.

We start this section by briefly reviewing the construction of predictive   intervals in a full parametric setup, followed by
reviewing the latest development of distribution-free conformal intervals for survival outcomes. We will also point out the difference and connection between  predictive intervals for a parameter and for predicting a future observation. Furthermore, we will discuss the limitations of existing conformal inference methods for censored survival outcome and propose new methods for such a setting.
\vspace{-3mm}
\section{Preliminaries}
\subsection{Traditional predictive intervals}

The prediction interval projects a range in which the outcome of a future patient will fall given a set of covariates, while a conventional confidence interval estimates the possible range of the  statistical parameter of the data, such as the population mean.
Suppose we have a training sample with independent and identically distributed  (iid) data $(Y_i,{\pmb X}_i), i=1,2,...,n$, where $Y_i$ is a response variable and ${\pmb X}_i$ is a $p$-dimensional vector. It is of interest to predict a future response $Y_{n+1}$ given a new set of covariates ${\pmb X}_{n+1}$.

 In the traditional prediction confidence approach, one may first postulate a parametric conditional density model $f(y|{\pmb X=\pmb x}; {\pmb\theta})$ in a regression model.
 The unknown $p$-dimension parameter ${\pmb\theta}$ associated with corresponding ${\pmb X}$ can be estimated by using the training samples. A  predictive set $C({\pmb x})$ can be obtained such that
 \begin{equation}\label{cond1}
 P(Y_{n+1}\in C({\pmb x})|{\pmb X}_{n+1}=\pmb x)
 \geq 1-\alpha \end{equation}
 approximately for a given mis-coverage probability of $\alpha$. 
 The methods have been  studied in   literature, for example,  \cite{dunsmore1980statistical} and \cite{geisser1993introduction}.

The methods based on the plug-in estimate $\widehat{\pmb \theta}$ have been criticized for ignoring the uncertainty of $\widehat{\pmb \theta}$ in  
the estimated confidence interval. Specifically, one needs to consider both the variation of the regression parameter estimate and the variation of the error term of the future observation. 
When sample size is small or modest, some modified versions of the methods were proposed to improve the coverage probabilities and robustness to model misspecification

\subsection{Conformal predictive interval}
\citet{vovk2005algorithmic} and \citet{vovk2009line} introduced a seminal concept of conformal prediction, 
 which recently has been utilised in statistical inference with novel machine learning algorithms \citep{lei2014distribution, lei2018distribution}. 
 The conformal prediction considers data uncertainty from a working model, which may not be correctly specified,  and constructs a statistically valid prediction interval for a future outcome $Y_{n+1}$ given an input set of covariate $\pmb{X}_{n+1}$ and training data. Using the same data notation as above, the conformal predictive interval is defined as 
\begin{equation}\label{conf1}
P(Y_{n+1}\in C({\pmb X}_{n+1}))\ge 1-\alpha
\end{equation}
where the probability is with respect to the joint distribution of $(Y_{n+1},{\pmb X}_{n+1})$ rather than to the conditional distribution, $P(Y_{n+1}\in C({\pmb x})|{\pmb X}_{n+1}=\pmb x)$ 
when constructing a traditional   predictive interval in (\ref{cond1}). 

One primary advantage of conformal predictive interval is that the conditional density of $Y$ given ${\pmb X}={\pmb x}$, $f(y|{\pmb x};{\pmb\theta})$  serves as a ``working model" for the relationship between the outcome and the covariates. A working model closer to the truth  will lead to a narrower  predictive interval and provide more informative prediction than otherwise. Regardless, the overall coverage probability of the conforma intervals remains at the specified level asymptotically. 

Several differences of the conformal  predictive interval over the traditional predictive confidence interval are worth noting. 
The conformal  predictive interval takes into consideration the variation of ${\pmb X}$ and $Y$. Thus, the interval is generally wider than the
 predictive confidence  interval, which only involves the variation of $Y$ for a fixed value of ${\pmb X}$.
Moreover, the conformal  predictive interval does not require the model $f(y|{\pmb x};{\pmb \theta})$ to
be correct.
For complex data structures with high dimension of ${\pmb X}$, 
conformal inference offers an effective approach to predicting outcome $Y$ by leveraging cutting-edge machine learning techniques.

\subsection{Conformal inference for survival data}

Despite the success on developing conformal intervals for uncensored data, there are limited works in conformal inference for general right-censored  data.
\cite{candes2023conformalized} recently proposed a novel method to estimate a covariates-dependent lower conformal prediction bound for  survival data subject to Type-I censoring.  
Consider iid observed data as
$Y_i=\min ({T}_i,C_i), \delta_i=I({T}_i\leq C_i), {\pmb X_i}, i=1,2,....,n,$
where ${T}_i, C_i$ are, respectively,  the  failure time and censoring time for individual $i$, $\delta_i$ is the censoring indicator, 
and {$\pmb X_i$} is a $p \times 1 $ vector of covariates.
\cite{candes2023conformalized} noted that in the presence of right censoring, 
\[
1-\alpha =P(T\geq q_{\alpha}({\pmb x}))
\geq P({T}\geq q_{\alpha}({\pmb x}))P(C\geq q_{\alpha}({\pmb x}))=P(Y\geq q_{\alpha}({\pmb x})) \] 
thus $P(Y\geq q_{\alpha}({\pmb x})) \le 1-\alpha$, 
where $q_{\alpha}(\pmb x)$ is a marginally calibrated lower probability bound of uncensored $T$ given $\pmb x$ at significance level $1-\alpha$. 
If the conformal  predictive interval of $T$ is constructed based on the observed  $Y=\min({T},C)$, it would be too conservative, especially when censoring occurred early.  Instead, they introduced an artificial truncation at a specified value $c_0$ by considering only training data samples with their censoring times $C\ge c_0$.
Under the assumption that $T$ and $C$ are independent given covariate ${\pmb x}$, 
$
P_{({\pmb X},\min(Y,c_0))|C\geq c_0}=P_{{\pmb X}|C\geq c_0}P_{\min(T,c_0)|{\pmb X}}.$
On the other hand,
$P_{({\pmb X},\min(T,c_0))}=P_{\pmb X} P_{\min(T,c_0)|{\pmb X}}.$
Thus, the distribution of $({\pmb X}, \min(T,c_0))$ in the subpopulation after right truncation and that of the whole population satisfies the so called covariate distributional shift \citep{Tibshirani2019}. 
 While it is novel to utilise artificial truncation to the training data and to apply the method of conformal prediction under covariate shift \citep{Tibshirani2019, lei2021conformal}, this approach is only applicable when the censoring time $C$ is fully observed (Type-I censoring) for all individuals  to construct the subpopulation from the training data. 
Still under the framework of artificial truncation, \cite{gui2022conformalized}  proposed a covariate-adaptive approach to finding the cutoff value of $c_0$  to improve the overly conservative bounds.

In medical research, we encounter more general censored survival data, where the censoring times are not  available for individuals whose failure event occurred before the random censoring event. Moreover, after excluding data observations with $C< c_0$ in \cite{candes2023conformalized},  
the constructed conformal predictive   interval is based on the truncated outcome $\min(T,c_0)$ rather than the true survival time $T$. 
Thus, the interpretation of the estimator is different.  Also, the choice of $c_0$ is tricky. A large value of $c_0$ results in a smaller subpopulation 
for training data, suffering more statistical efficiency loss. A small value of $c_0$ produces a less informative conformal   interval. 
More importantly, as we will show in the motivating data analysis, two-sided predictive intervals provide more critical information for medical decision making in practice than a one-sided lower bound.

In this paper, we propose an approach to constructing two-sided as well as one-sided conformal predictive   intervals for survival data with general censoring distributions.  
We utilize the nonparametric and semiparametric bootstrap methods by \cite{efron1979bootstrap}  to directly approximate the pivot used for the conformal predictive intervals. We further extend the algorithm to several practical directions in medical research.

\section{Bootstrap conformal predictive intervals}

Denote the joint cumulative distribution function (CDF) of the failure time and covariate as $F(t,{\pmb x})$, and 
the conditional survival function of the failure time and censoring time as 
$ S(t|{\pmb x})=P(T>t|{\pmb x}), G(t|{\pmb x})=P(C>t|{\pmb x}).$
The key to the conformal inference is that one must find a nonparametric estimation of the joint distribution $F(t,{\pmb x})$. 
The probability of observing a subject that failed at time $T=t$ with covariate $\pmb{X=x}$ (i.e., $\delta=1$) is given by 
\begin{equation}\label{equ1}
P(T=t,{\pmb X=x}|\delta=1) \sim \frac{G(t|{\pmb x} )dF(t,{\pmb x})}{P(C>T)}.
\end{equation}

We propose a new method to handle the general censored survival data when the shift function depends on the outcome. 
We first consider $G(t|\pmb x)=G(t)$, i.e., the censoring variable is independent of the covariates, and 
will present the general case
where censoring distribution depends on ${\pmb X}$ later. 
The joint CDF of $(T, \pmb{X})$ can be consistently estimated by its empirical distribution function with an inverse weighted censoring distribution using the training data, 
\begin{equation}\label{joint}
 \widehat{F}(t,{\pmb x})=\frac{1}{n}\sum_{i=1}^{n} \frac{\delta_iI(Y_i\leq t,{\pmb X_i}\leq {\pmb x})}{\widehat{G}(Y_i)}, 
\end{equation}
where $\widehat{G}(.)$ is estimated by the Kaplan-Meier method for censoring times.  Note that the marginal distribution of $T$, 
the estimate in Equation (\ref{joint}), is reduced to
$
\widehat{F}(t)=\frac{1}{n}\sum_{i=1}^{n}\frac{\delta_i I(Y_i\leq t)}{\widehat{G}(Y_i)},$
which is precisely the Kaplan-Meier estimate of failure time \citep{satten2001kaplan}.

To construct a conformal predictive interval,  it is common to utilize a working regression model for the failure time with given covariates. This model could be a parametric model or a semiparametric model. For the purpose of illustration,  we use the Cox model as a working model without loss of generality. Denote the conditional survival function as
\[ S(t| {\pmb x}; \pmb\theta)= \exp\{-\Lambda(t| {\pmb x})\}, ~~ \Lambda(t| {\pmb x})= \Lambda_0(t) \exp({{\pmb x}\pmb\beta}),\] 
where $\Lambda_0(t)$ is the baseline cumulative hazard function and ${\pmb\theta}=
({\pmb\beta},{\Lambda}$). 
We use the observed training data $(Y_i,\delta_i,{\pmb X_i}), i=1,2,....,n$ to
estimate the unknown parameters by the partial likelihood method and the Breslow estimate, denoted as $\widehat{\pmb\theta}=
(\widehat{\pmb\beta},\widehat{\Lambda}$). Let $({\pmb X}_{n+1},T_{n+1})$ be the testing data, where $X_{n+1}$ is given, but $T_{n+1}$ is to be predicted. 
We describe the step-by-step 
procedure to construct a conformal predictive interval for $T_{n+1}$:  
\begin{itemize} 
\item[]{\bf Step 1.} Generate the bootstrap re-sampling data,  ${\mathcal D}^*=(Y_i^*,\delta_i^*,{\pmb X_i}^*, i=1,2,...,n)$, from the original training data with equal probability and replacement.
 \item[]{\bf Step 2.} Obtain $\widehat{\pmb\theta}^* =  (\widehat{\pmb\beta}^*,\widehat{\Lambda}^*)$ under the working model with the bootstrap dataset,   ${\mathcal D}^*$. 
 \item[]{\bf Step 3.} Randomly sample one pair $(T^b, {\pmb X}^b)$ from the  estimated empirical joint CDF, $\widehat{F}(t,{\pmb x})$ in (\ref{joint}), which is estimated using the original training data.
\item[]{\bf Step 4.}
 Calculate $U^b=\exp\{-\widehat{\Lambda}^*(T^b)\exp({\pmb X}^b\widehat{\pmb\beta}^*)\}$.
\item[]{\bf Step 5.} After repeating {\bf Steps 3-4} for $B$ times,  denote $L$ and $R$ as the low $[B\alpha/2]$-th and upper $[B(1-\alpha/2)]$-th quantile of $(U^b, b=1, \cdots, B)$, respectively.
\item[]{\bf Step 6.} Construct the conformal interval by solving the following inequality for $T_{n+1}$: 
\[
L\leq \exp\{-\widehat{\Lambda}(T_{n+1})\exp({\pmb X}_{n+1}\widehat{\pmb\beta})\}\leq R,~~ \mbox{or}
\]
\[
\widehat{\Lambda}^{-1}\{-(\log R) e^{-{\pmb X}_{n+1}\widehat{\pmb\beta}}\}\leq T_{n+1}\leq \widehat{\Lambda}^{-1}\{-(\log L) e^{-{\pmb X}_{n+1}\widehat{\pmb \beta}}\}.
\]
\end{itemize}
The proposed procedure under a general working model is described in Algorithm 1, in which $\pmb{\theta}$ defines all unknown parameters of the working model.  In simulation studies, we assess the performance of the predictive interval under various working models
including log-normal and Weibull models. 


\begin{algorithm}
\caption{Conformal prediction with right-censored data}\label{alg:cap}
\begin{algorithmic}
\State \textbf{Input:} Training data $(Y_i, \delta_i, {\pmb X_i}), i=1, \cdots, n,$ miscoverage level  $\alpha$, number of bootstrap replicates $B$, working model $S(t\mid{\pmb x};\pmb{\theta})$, using the original training data estimate  $\pmb{\theta}$, $\widehat{\pmb\theta}$, and 
the empirical  estimator of the joint CDF of $(T, \bf{X})$, $\widehat{F}(t,{\pmb x})$.  Points $\mathcal{X}_{new}=\{{\pmb x}_{n+1}, {\pmb x}_{n+2}, \cdots\}$ at which to construct prediction intervals. 
\State\textbf{Output}: Prediction intervals at each element of $\mathcal{X}_{new}$
\vspace*{4pt}
\State\textbf{Step 1:} Generate the bootstrap data 
${\mathcal D}^*=(Y_i^*,\delta_i^*,{\pmb X_i}^*, i=1,2,...,n)$ from the original data with equal probability and replacement.
\State\textbf{Step 2:} Obtain $\widehat{\pmb\theta}^*$ under the working model  using ${\mathcal D}^*$.
\For{$b = 1,2, ..., B$ }
\State\textbf{Step 3:} Sample one pair $(T^b, {\pmb X}^b)$ from the joint distribution  $\widehat{F}(t,{\pmb x})$.
\State\textbf{Step 4:} Calculate $U^b=S(T^b \mid {\pmb X}^b;  \widehat{\pmb\theta}^*)$.
\EndFor
\State  Define $L$ and $R$ as the $[B\alpha/2]$-the and
$[B(1-\alpha/2)]$-th quantile among $U^b, b=1, \cdots, B$. 
\State  Return $\mathcal{C}_{conf}({\pmb x})=[S^{-1}(R\mid{\pmb x};  \widehat{\pmb\theta}), S^{-1}(L \mid {\pmb x};  \widehat{\pmb\theta}) ]$ for each $\pmb{x} \in \mathcal{X}_{new}$
\end{algorithmic}
\end{algorithm}

Note that  the conformal predictive interval under a semiparametric working model (e.g., Cox model) cannot be extrapolated beyond the largest observed failure time $\eta=\max\{Y_i, \delta_i=1, i=1,2,...,n\}$. 
In particular, when censoring is heavy in the training data, estimation of the predictive   interval, especially the upper bound can be inaccurate. 
 Instead, we can provide a more reliable predictive   interval for the minimum of failure time $T_{n+1}$ and $\eta$. 
By modifying {\bf Step 6}, the estimated conformal interval of $\min(T_{n+1},\eta)$ is 
$\widehat{\Lambda}^{-1}\{-(\log R)e^{-{\pmb X}_{n+1}\widehat{\pmb\beta}}\}\leq T_{n+1}\leq \min(\eta, \widehat{\Lambda}^{-1}\{-(\log L)e^{-{\pmb X}_{n+1}\widehat{\pmb \beta}}\}.$ 

Note that the censoring distribution, $G(t|{\pmb x})$ possibly depends on the failure time via the covariates, playing a role of shifting that results in a biased sampling version of $F(t,{\pmb x})$.   When predicting a failure time $T$ with ${\pmb X}$ but the observed training samples have censored outcomes, the covariate shift conformal inference method developed by \cite{Tibshirani2019} is not directly applicable to the setting with an outcome-dependent shift function. This issue is often referred to as prior probability shift or label shift in machine learning \citep{dockes2021preventing}. The resemblance between the shift function and weight function in the likelihood-adjusting non-ignorable missing data was described in following {Remark}:

  Conformal inference in survival analysis involves tackling a more intricate shift problem, wherein the shift function is contingent on both failure time $T$ and covariate $\pmb X$. This bears a resemblance to non-ignorable missing data problems, where the missing probability function $w(t,\pmb x)$ (in our context, the shift function) is dependent on both the outcome variable and covariate. Unlike the missing-at-random problem, where the outcome likelihood and missing data likelihood are separable in the likelihood representation, the non-ignorable missing data problem intertwines the outcome density and likelihood of the missing data. 

Consider a general shift problem where the shift function 
$w(t,\pmb x)$  depends on both $\pmb X$ and $T$.
Suppose the joint distribution of observed training 
data $(T_i,\pmb X_i)$, $i=1,2,...,n$ follows 
$\frac{w(t,\pmb x)dF(t, \pmb x)}{\Delta},$ where $\Delta$ is the normalized term associated with the shift function. The test data $(T_{n+1},\pmb X_{n+1})$ follow a joint distribution $F(t,\pmb x)$. 
We can show that if $w(t,\pmb x)$ is correctly specified, 
the inverse weighted estimator 
\[\psi_n(u)=\frac{\sum_{i=1}^nw^{-1}(T_i,\pmb X_i)
I(S(T_i|\pmb X_i)\leq u)}{\sum_{i=1}^nw^{-1}(T_i,\pmb X_i)}\] consistently estimates $P(S(T_{n+1}|\pmb X_{n+1})\leq u).$
On the other hand, if the  shift function depends on the covariate alone, i.e., $w(t,\pmb x)=w(\pmb x)$ but may possibly be misspecified,  \[
\psi_n(u) \rightarrow 
E[w^{-1}(\pmb X)w_0(\pmb X)I(S(T|\pmb X)\leq u)]
E[w^{-1}(\pmb X)w_0(\pmb X)]^{-1},
\]
where $w_0(\pmb X)$ is the true shift function. 
If the conditional survival  function, $S(t|\pmb x)$, is correctly specified  given $\pmb X=\pmb x$, then $S(T|\pmb X)$ has a uniform distribution in $[0,1]$. Thus
$
E[I(S(T|X)\leq u)|X]=u.$
Subsequently, 
\[
E[w^{-1}(\pmb X)w_0(X)I(S(T|\pmb X)\leq u)]
E[w^{-1}(\pmb X)w_0(\pmb X)]^{-1} 
=u \]
and $
P(S(T_{n+1}|\pmb X_{n+1})\leq u)=u$
 regardless of if $w(\pmb x)=w_0(\pmb x)$  or not. 
Therefore, the double robustness can be achieved if and only if the shift function is independent of outcome $T$. This is the so called covariate shift problem. 
In the general setting of survival analyses, the shift function $w(t, \pmb x)$ is a function of censoring distribution $G(t\mid \pmb x)$, which always depends on $T$ (and possibly on $\pmb X$ too).

The following theoretical properties of the proposed conformal prediction interval  ensure the average coverage probability to be valid asymptotically. Without loss of generality, suppose the first $n_1$ observations are observed failures ($\delta_1=\ldots=\delta_{n_1}=1$), and let $n_1=\sum_{i=1}^n\delta_i$, and $\widehat{\Delta}=\sum_{i=1}^{n_1}\widehat{G}^{-1}(Y_i)$. We first establish the asymptotic behaviour  of the estimated CDF of the failure time and covariate, i.e., 
\begin{equation}\label{JE}
\widetilde{F}(t, {\pmb x})=\frac{\widehat{F}(t, {\pmb x})}{\widehat{F}(\infty,\infty)}=\sum_{i=1}^{n_1}\frac{\widehat{\Delta}}{\widehat{G}(Y_i)}I(Y_i\leq t, {\pmb X}_i\leq {\pmb x}).
\end{equation}

\noindent{\bf Theorem 1.} Under regularity conditions (C.1)-(C.4) in the Appendix,  $\widetilde{F}(t,{\pmb x})$ is a consistent estimate of 
$F(t,{\pmb x})$, and 
$\sqrt{n}\{\widetilde{F}(t,{\pmb x})-F(t,{\pmb x})\}$ converges in distribution to 
a normal distribution with mean 0 and variance $\sigma(t,{\pmb x})$, where  $\sigma(t,{\pmb x})$ is defined in the Appendix. 

\noindent{\bf Theorem 2.} Under  regularity conditions (C.4)-(C.9) in the Appendix, the bootstrap conformal  predictive interval is valid, ensuring correct coverage of $100(1-\alpha)\%$ asymptotically.

\section{Several extensions for survival conformal inference} 

\subsection{Covariate-dependent censoring distribution}\label{cencov}
In medical research, the large training datasets often include multi-site patient samples with site-specific  or the patients' covariate-specific censoring distribution. 
The method developed above can be generalized to the case when $G(t|\pmb{x})$ depends on covariate $\pmb{x}$ 
in (\ref{joint}). 

When the censoring distribution is study-site dependent, we can estimate $G_k, k=1,2,...,K$ for each study site separately,using the Kaplan-Meier estimate as  $\widehat{G}_k(t)$, where $K$
is the number of study sites. Without loss of generality, $F(t,\pmb{x})$ remains the same across sites, and $P(C\geq t|S=k)=G_k(t)$.
The joint CDF of $(T,\pmb X)$ can be consistently estimated by the inverse weighted empirical function, 
\begin{equation}\label{joint2}
 \widehat{F}(t,{\pmb x})=\frac{1}{K} \sum_{k=1}^K \sum_{i=1}^{n_k} \frac{\delta_{ki} I(Y_{ki}\leq t,{\pmb X}_{ki}\leq {\pmb x})}{n_k\widehat{G}_k(Y_{ki})}, 
\end{equation}
where the observed data are denoted as $(Y_{ki},\delta_{ki},{\pmb X}_{ki}),$ $ i=1, \cdots, n_k,$ and $k=1,\cdots, K$.

When the censoring distribution depends on a set of covariates rather than a categorical variable such as the study site, one could postulate a parametric or semiparametric regression model for $C$ given ${\pmb X}$.
For example, we assume the Cox model for $G(t|{\pmb X})=G(t|{\pmb X}; \pmb{\gamma})$, and 
 the unknown parameters $\pmb{\gamma}$ can be estimated by maximising the likelihood, \[
\prod_{i=1}^n[G(Y_i|{\pmb X}_i; \pmb{\gamma})]^{\delta_i}[dG(Y_i|{\pmb X}_i; \pmb{\gamma})]^{1-\delta_i}. \]
The joint CDF of $(T, \pmb{X})$ can be consistently estimated by its empirical distribution function with the inverse weighted censoring distribution $\widehat{G}(t|{\pmb X}; \widehat{\pmb{\gamma}})$,  
\begin{equation}\label{joint1}
 \widehat{F}(t,{\pmb x})=\frac{1}{n}\sum_{i=1}^{n}\frac{\delta_iI(Y_i\leq t,{\pmb X_i}\leq {\pmb x})}{\widehat{G}(Y_i|{\pmb X}_i; \widehat{\pmb{\gamma}})}. 
\end{equation}
 It is possible to estimate the covariate dependent censoring distribution nonparametrically by using the kernel method if the dimension of ${\pmb X}$ is low or $C$ depends on ${\pmb X} $ through only a few components of ${\pmb X}$. However, this could be computationally intensive. 

\vspace{3mm}

\noindent{\bf Theorem 3.} Under  regularity conditions (C.4)-(C.9) in the Appendix, the bootstrap conformal  predictive interval with covariate-dependent censoring is valid, ensuring correct coverage of $100(1-\alpha)\%$ asymptotically.

\subsection{Conformal interval for remaining survival time}

Another important practical clinical question is to predict a patient's survival time, given that the patient survived for $c_L$ (e.g., 2 years) after her initial diagnosis considering her tumour characteristics and treatment received.   We can construct a predictive conformal interval for her remaining life expectancy given that the individual survived $c_L$ years with covariate $\pmb X_{n+1}$. We are interested in predicting %
the remaining lifespan, $T_{n+1}- c_L$. 
Note that
\[
P(T_{n+1}>t|T_{n+1}>c_L, \pmb X_{n+1})=
\frac{S(t|\pmb X_{n+1})}{S(c_L|\pmb X_{n+1})}, ~~ t>c_L>0.
\]
Interestingly, the above equation shows a covariate shift for the distribution of a subpopulation of patients who survived $c_L$ years from that of the original whole population in the training data. We can modify the methods in the above Section to incorporate the intermediate information as follows: 
\begin{itemize} 
\item[]{\bf Step 1.}
 Generate the bootstrap re-sampling data,  ${\mathcal D}^*=(Y_i^*,\delta_i^*,\pmb{X}^*_i, i=1,2,...,n)$,  from the original training data with equal probability and replacement.

 \item[]{\bf Step 2.} Obtain $\widehat{\pmb\theta}^*=(\widehat{\pmb\beta}^*,\widehat{\Lambda}^*)$ by fitting the working regression model using the bootstrap data, i.e.,  ${\mathcal D}^*$.

\item[]{\bf Step 3.} Randomly sample pair $(T^b, {\pmb X}^b)$ from the conditional CDF, $\widehat{F}(t,{\pmb x}|t>c_L)$. This can be achieved by
sampling pairs from $\widehat{F}(t,{\pmb x})$ but only keeping those $(T^b,{\pmb X}^b)$ with $T^b>c_L$.

\item[]{\bf Step 4.}
 Calculate 
 $U^b=S(T^b|{\pmb X}^b;\widehat{\pmb\theta}^*)/S(c_L|{\pmb X}^b; \widehat{\pmb\theta}^*), T^b\geq c_L$. 

\item[]{\bf Step 5.} After repeating {\bf Steps 3-4} for $B$ times,  obtain $L$ and $R$ as the low $[B\alpha/2]$-th and upper $[B(1-\alpha/2)]$-th quantile of $(U^b, b=1, \cdots, B)$, respectively.

\item[]{\bf Step 6.} Construct the conformal interval by solving the following inequality, \\
$
L\leq \frac{S(T_{n+1}|{\pmb X}_{n+1};\widehat{\pmb\theta})}{S(c_L|{\pmb X}_{n+1}; \widehat{\pmb\theta} )}\leq R $
or equivalenty
$ S(c_L|{\pmb X}_{n+1};\widehat{\pmb\theta}) L\leq S(T_{n+1}|{\pmb X}_{n+1};\widehat{\pmb\theta})\leq  S(c_L|X_{n+1};\widehat{\pmb\theta})R. $
The $100(1-\alpha)\%$ conformal predictive  interval is
\begin{eqnarray*}
 S^{-1}\{S(c_L|{\pmb X}_{n+1};\widehat{\pmb\theta})R|{\pmb X}_{n+1}\}\leq  T_{n+1}
 \leq   S^{-1}\{S(c_L|{\pmb X}_{n+1}; \widehat{\pmb\theta}) L|{\pmb X}_{n+1}\}.
\end{eqnarray*}
\end{itemize}

\subsection{Split conformal prediction to survival outcomes}
In contrast to the split conformal prediction method proposed for uncensored outcomes in \cite{lei2018distribution}, we describe a different split approach to predicting a conformal interval for failure times. The motivation for this extension is to validate the performance (i.e., average coverage probability) of the estimated predictive   intervals given covariates in a {\it real data analysis} setting when we do not observe the failure times for censored subjects. 
We randomly split the original data into training and testing folds. 
For partially missing outcome data due to censoring, one cannot validate if a censored observation in the testing fold belongs to the predictive interval derived from the training fold. In contrast, we modified the fitting algorithm for the training data proposed in the above section to align with the purpose of assessing the coverage probability of observed uncensored observations in the testing data.

We randomly split the original data into a training fold, denoted by $
{\mathcal I}_1=\{(Y_i,\delta_i,X_i), i=1,2,....,n\}$, 
and a testing fold, denoted by ${\mathcal I}_2=\{(Y_i,\delta_i,X_i), i=n+1,2,....,2n\}$. For  simplicity of notation, we used equal size for the two folds, but unequal size splits may be used.
For model validation purposes, we are interested in predicting an uncensored failure sample in the testing fold, denoted as $Y_{n+i}$ with $\delta_{n+i}=1, i=1,2,...,n$. Without loss of generality, we denote the first $m_1$ ($m_1 \le n$) samples with uncensored failure times in the training fold as $(Y_1,X_{1},\delta_{1}=1),...,(Y_{m_1},X_{m_1},\delta_{m_1}=1)$, and the first $m_2$ ($m_2 \le n$)
samples with uncensored failure times in the testing fold as $(Y_{n+1},X_{n+1},\delta_{n+1}=1),...,(Y_{n+m_2},X_{n+m_2},\delta_{n+m_2}=1)$.
For the  purpose of validation in real data analysis, we describe the following analysis steps: 
\vspace{-.1in}
\begin{itemize} 
\item[]{\bf Step 1.}
 Randomly split the original data into a training ${\mathcal I}_1$ and testing ${\mathcal I}_2$ fold.
 
\item[]{\bf Step 2.} Use the training fold ${\mathcal I}_1$, generate the bootstrap re-sampling data,  ${\mathcal D}^*=(Y_i^*,\delta_i^*,{\pmb X_i}^*, \\i=1,2,...,n)$,  with equal probability and replacement. 

\item[]{\bf Step 3.} Obtain $\widehat{\pmb\theta}^*$
by fitting the working regression model using ${\mathcal D}^*$ from ${\mathcal I}_1$.

\item[]{\bf Step 4.} Randomly 
sample $(T^b,{\pmb X}^b)$ from uncensored observations of  ${\mathcal I}_1$ with equal probability; and 
calculate 
$U^b=S(T^b|{\pmb X}^b;\widehat{\pmb\theta}^*)$. 

\item[]{\bf Step 5.} Repeat {\bf Step 4} $B$ times, obtain the lower $[B\alpha/2]$-th and upper $[B(1-\alpha)]$th quantile of $U^b$ as $L$ and $R$, respectively. 

 \item[]{\bf Step 6.} For any given uncensored pair $(Y_{n+j}, {\pmb X}_{n+j})$ for $j=1,2,...,m_2$ in the testing fold, solve $Y_{n+j}$ from 
$L\leq S(Y_{n+j}|{\pmb X}_{n+j};\widehat{\pmb\theta})\leq R$ to obtain
the predictive   interval $\mathcal{C}_{conf}({\pmb X}_{n+j})=
[S^{-1}(R; {\pmb X}_{n+j},  \widehat{\pmb\theta}), S^{-1}(L; {\pmb X}_{n+j},  \widehat{\pmb\theta}) ]$.

\item[]{\bf Step 7.} In the testing fold uncensored failure time $Y_{n+j}, j=1,2,..,m_2$ are observed, summarize the proportion that 
$Y_{n+j} \in \mathcal{C}_{conf}({\pmb X}_{n+j})$  $j=1,2,...,m_2$. 
\end{itemize}

\vspace{-.1in}
For this particular extension to assess the performance in real data applications, the focus here is different from the previous sections. 
We just want to estimate the overall coverage proportion for observed failure times using the constructed intervals from {\bf Steps 5-6} to see if it is close to the specified coverage probability $(1-\alpha)$. 
One may also consider multiple splits by repeating {\bf Steps 1-6} to obtain a more robust estimate of the coverage probability. 

\section{Simulation studies}

We conducted extensive simulation studies to evaluate the performance of the proposed methods using various working regression models under different scenarios. We assume the covariate vector included four components $\pmb X=(X_1,X_2, X_3,X_4)$, where $X_1$ and $X_3$ were binary variables following a Bernoulli distribution with probability 0.5, $X_2$ and $X_4$ are continuous variables following a Uniform distribution on $U(-5,5)$, and a standard normal distribution, respectively. The failure time $T$ was generated either from the Weibull or log-normal distribution.  The censoring time C was generated independently from a uniform distribution $U(0, \tau_c)$, and $\tau_c$  was chosen to get the desired censoring rates of 15\% or 50\%. For each scenario, there were 2000 repetitions with a training sample size of 2000 and a testing sample size of $n_{test}=2000.$

Using the proposed conformal inference method, a working regression model was needed. Though the working model does not have to be correct, we want to evaluate how the selection of a working model might impact on efficiency and accuracy under different underlying failure and censoring distributions. Simulation results of the proposed conformal predictive intervals (CPI) will be summarized under three working models denoted as follows:
\begin{enumerate}
    \item CPI-log-normal: $S(t|{\pmb x}; {\pmb\theta})$ $\sim$ log-normal working model  
    \item CPI-Weibull: $S(t|{\pmb x}; {\pmb\theta})$ $\sim$ Weibull working model
    \item CPI-Cox: $S(t|{\pmb x}; {\pmb\theta})$ $\sim$ Cox working model
\end{enumerate}
For comparison, the conventional conditional approaches (4)-(6) were also considered. These conditional approaches rely on the correct specification of the model. If the model was correctly specified, the conditional approaches would perform well. If a wrong model was used, the method may not be valid. We also compare the proposed method with two distribution-free approaches, a random forest model (7) using an ensemble of decision trees and a marginal Kaplan-Meier (K-M) estimate ignoring covariates information (8). 
\begin{enumerate}
\setcounter{enumi}{3}
    \item Conditional log-normal model
    \item Conditional Weibull model
    \item Conditional Cox model
    \item Random forest model
    \item Marginal K-M estimation without using covariates
\end{enumerate}
For each of the conditional approaches (4-6), we first estimated $\widehat{\pmb\theta}$ using the training data under the respective model assumed,  and then for a new observation in the testing dataset, we obtain $\widehat{S}(t_{n+1}|{\pmb x_{n+1}}; \widehat{\pmb\theta})$. The K-M estimate served as a reference without using covariate information in the testing dataset. 
For the one-sided interval, the lower bound was calculated as the $\alpha$-th quantile from the estimated model; for the two-sided interval, the prediction interval was estimated from the ${\alpha}/{2}$-th and  $(1-{\alpha}/{2})$-th quantiles of the conformal scores.

We evaluated the coverage rate of the prediction interval as $\sum_{i=1}^{n_{test}} I\{T_i \ge \widehat L(X_i)\}/n_{test} $ for the one-sided bound and $\sum_{i=1}^{n_{test}} I\{\widehat L(X_i) \le T_i  \le  \widehat U(X_i)\}/n_{test} $ for the two-sided interval. Besides coverage probability, we also calculated the mean and standard deviation of the lengths of the prediction intervals as $ \sum_{i=1}^{n_{test}} \{ \widehat U(X_i) - \widehat L(X_i)\}/n_{test} $ for the two-sided interval. Wider prediction intervals indicated less efficient estimates with higher uncertainty.  
A smaller lower bound value represented higher uncertainty for the one-sided interval.

\begin{figure}
\centering
\includegraphics[scale=0.32]{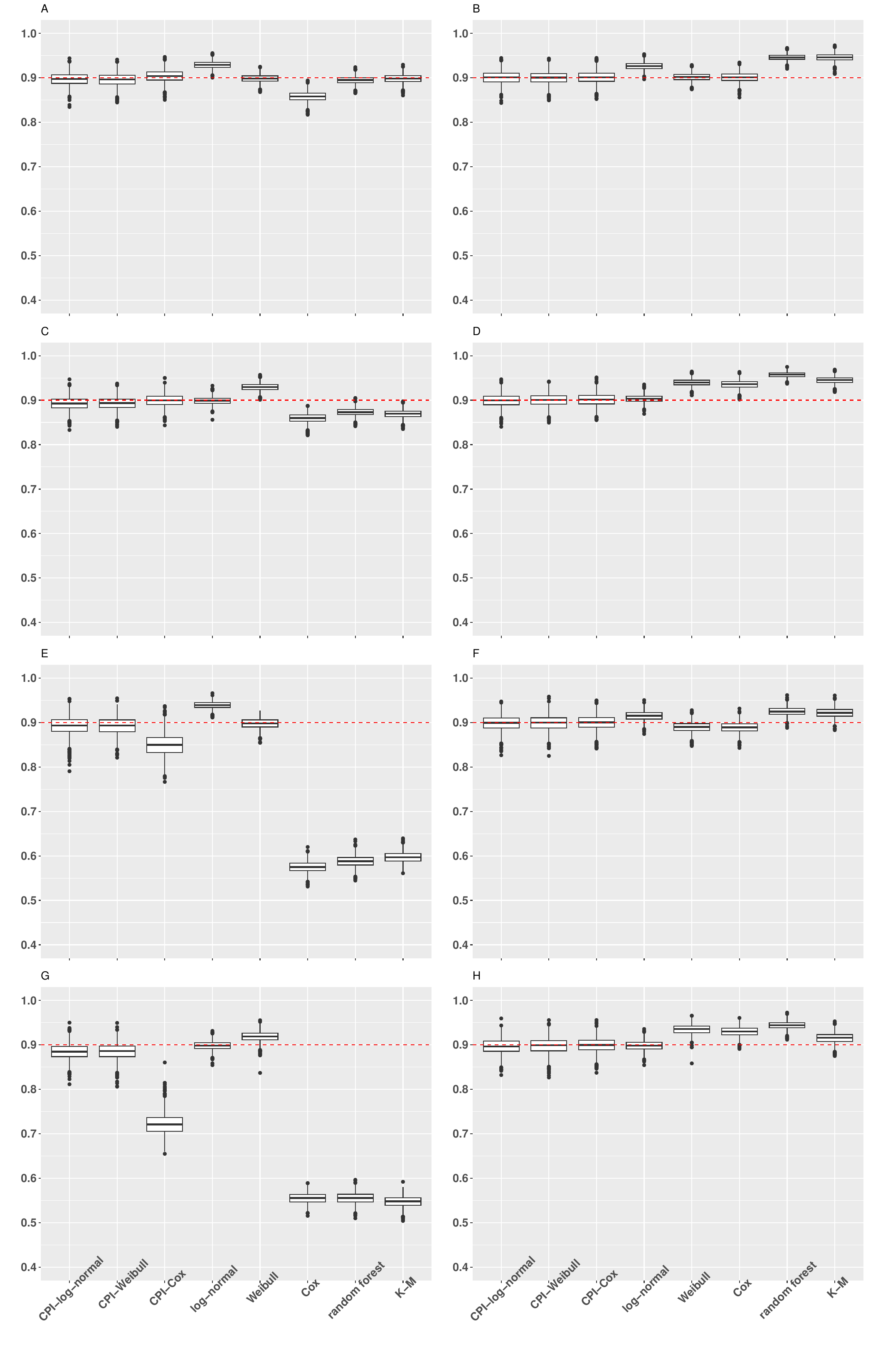}
\caption{Two-sided empirical 90\% coverage of the failure times where data were from (A) Weibull model, censoring rate = 15\%; (B) Weibull, min(failure times, $\eta$), censoring rate = 15\%; (C) log-normal, censoring rate = 15\%; (D) log-normal, min(failure times, $\eta$),  censoring rate = 15\%; (E) Weibull model, censoring rate = 50\%; (F) Weibull, min(failure times, $\eta$), censoring rate = 50\%; (G) log-normal, censoring rate = 50\%; (H) log-normal, min(failure times, $\eta$),  censoring rate = 50\%.}
\label{fig1}
\end{figure}

Table \ref{tab1}  and Figure \ref{fig1} 
summarized the results from the two-sided predictive intervals under 15\% and 50\% censoring rates.  The proposed approaches under different working models (CPI-log-normal, CPI-Weibull, CPI-Cox) all performed well, maintaining the specified coverage probability while having a reasonable prediction interval length when the censoring rate is low (15\%). The correct specification of working models was not necessary; it did not impact the coverage rate nor the length of the prediction interval much, but smaller variations for the interval lengths are observed when the working model is correctly specified.
It is interesting to note that CPI-Cox achieved the specified coverage probability. The average length of the CPIs under the Cox model are much narrower and even better than these of CPI-Weibull and CPI-log-normal under the correct model assumptions. This is likely because the maximum support of CPI-Cox is at the largest observed failure time; the estimated CPI and its length are thus robust to the extreme failure times due to long-tailed parametric distributions. 
As expected, the conditional estimators are sensitive to the model assumption; they could lead to over or under-coverage if the working model is different from the true underlying distribution. When failure times follow log-normal distribution, the nonparametric K-M, random forest, and conditional Cox model were unable to maintain the specified coverage rate, 90\%, for the two-sided problem. 

\begin{table}
\centering
  \caption{Empirical two-sided 90\% coverage of the failure times.   \label{tab1}}
\begin{tabular}{lllcccccc}

\toprule
           &   &       &     \multicolumn{3}{@{}c@{}}{Failure times}            & \multicolumn{3}{@{}c@{}}{min(Failure times, $\eta$)} \\
              \cline{4-6}\cline{7-9}%

    Censoring &Distribution & Method & Cov   & Length & SD    & Cov   & Length & SD \\\midrule
       15\%  & Weibull &  CPI-log-normal & 0.90  & 21.35 & 1.67  & 0.90  & 13.82 & 1.02 \\
          &       & CPI-Weibull & 0.90  & 20.84 & 1.55  & 0.90  & 13.53 & 0.97 \\
          &       & CPI-Cox & 0.90  & 11.98 & 0.51  & 0.90  & 10.41 & 0.52 \\
          &       & log-normal & 0.93  & 31.08 & 2.35  & 0.93  & 20.11 & 1.44 \\
          &       & Weibull & 0.90  & 21.52 & 1.35  & 0.90  & 13.97 & 0.84 \\
          &       & Cox   & 0.86  & 12.07 & 0.50  & 0.90  & 10.55 & 0.54 \\
          &       & random forest & 0.90  & 14.97 & 0.81  & 0.95  & 13.57 & 0.81 \\
          &       & K-M   & 0.90  & 46.40 & 3.58  & 0.95  & 46.40 & 3.58 \\
          &       &       &       &       &       &       &       &  \\
          & log-normal &  CPI-log-normal & 0.89  & 152.82 & 21.00 & 0.90  & 50.56 & 6.20 \\
          &       & CPI-Weibull & 0.89  & 147.48 & 20.93 & 0.90  & 49.12 & 6.05 \\
          &       & CPI-Cox & 0.90  & 29.59 & 1.43  & 0.90  & 24.70 & 1.47 \\
          &       & log-normal & 0.90  & 169.58 & 19.53 & 0.90  & 56.10 & 5.38 \\
          &       & Weibull & 0.93  & 152.98 & 18.66 & 0.94  & 50.95 & 5.10 \\
          &       & Cox   & 0.86  & 31.76 & 1.42  & 0.94  & 25.67 & 1.52 \\
          &       & random forest & 0.87  & 37.91 & 2.45  & 0.96  & 32.64 & 2.35 \\
          &       & K-M   & 0.87  & 119.21 & 0.84  & 0.95  & 119.21 & 0.84 \\
          &       &       &       &       &       &       &       &  \\
    50\%  & Weibull &  CPI-log-normal & 0.89  & 50.88 & 6.28  & 0.90  & 7.53  & 0.79 \\
          &       & CPI-Weibull & 0.89  & 50.39 & 5.46  & 0.90  & 7.49  & 0.74 \\
          &       & CPI-Cox & 0.85  & 3.88  & 0.20  & 0.90  & 3.60  & 0.15 \\
          &       & log-normal & 0.94  & 87.01 & 10.61 & 0.92  & 12.88 & 1.34 \\
          &       & Weibull & 0.90  & 54.46 & 5.26  & 0.89  & 8.09  & 0.71 \\
          &       & Cox   & 0.58  & 3.55  & 0.11  & 0.89  & 3.62  & 0.13 \\
          &       & random forest & 0.59  & 4.13  & 0.23  & 0.93  & 4.26  & 0.20 \\
          &       & K-M   & 0.60  & 8.90  & 0.01  & 0.92  & 8.90  & 0.01 \\
          &       &       &       &       &       &       &       &  \\
          & log-normal &  CPI-log-normal & 0.89  & 137.29 & 24.62 & 0.90  & 3.06  & 0.44 \\
          &       & CPI-Weibull & 0.89  & 136.36 & 34.71 & 0.90  & 3.05  & 0.46 \\
          &       & CPI-Cox & 0.72  & 1.00  & 0.05  & 0.90  & 1.09  & 0.06 \\
          &       & log-normal & 0.90  & 171.29 & 26.89 & 0.90  & 3.81  & 0.43 \\
          &       & Weibull & 0.92  & 151.00 & 36.14 & 0.94  & 3.38  & 0.40 \\
          &       & Cox   & 0.56  & 1.11  & 0.04  & 0.93  & 1.16  & 0.06 \\
          &       & random forest & 0.56  & 1.20  & 0.08  & 0.94  & 1.37  & 0.08 \\
          &       & K-M   & 0.55  & 2.99  & 0.00  & 0.92  & 2.99  & 0.00 \\
\bottomrule
    \end{tabular}%
\end{table}%

The simulation results with heavy censoring, 50\%, showed that the proposed CPI with parametric working models performed well, achieving 89\% coverage probability given the desired 90\% coverage probability, whereas the CPI-Cox approach 
suffered an under-coverage issue because the survival distribution was not properly estimated beyond the maximum observed failure time when the support of the censoring distribution is shorter than that of the survival distribution.  
Note that the nonparametric confidence intervals using the K-M $F_{KM}^{-1}(\alpha/2)$ and $F_{KM}^{-1}(1-\alpha/2)$, random forest, or the conditional Cox model suffered even worse coverage probabilities than CPI-Cox. If the model is misspecified as the log-normal model when failure times were generated from the Weibull distribution, we observed a wide prediction interval length with an overestimated coverage rate. Conversely, when data are generated from the log-normal distribution, the conditional log-normal approach can achieve coverage probability as expected with a smaller variation in the prediction interval length. In the presented scenarios, CPI-log-normal and CPI-Weibull achieved shorter average interval lengths compared to the conditional CIs under even correctly specified distribution (e.g., 136 or 137 compared with 151 under log-normal).  

In summary, if censoring is light in the training dataset, conformal predictive interval with the Cox working model is recommended for its estimation accuracy (much narrower CI) and robustness compared with the CPIs under parametric working models. In contrast, if censoring is heavy, the upper bound of CPI can be inaccurate for CPI with a semiparametric working model or nonparametric methods. In this case, 
using a parametric working model over the semiparametric model for CPI is preferable to ensure a reasonable coverage probability. 
Alternatively, we can provide a more reliable predictive   interval for the minimum of failure time $T^*$ and $\eta$, which is the maximum of observed failure times in the training data, regardless of censoring rate, as shown in the second column of Figure 1. Note that the length of CPI was much narrower for this estimate compared with its counterpart with a parametric working model. For other approaches, we also observed improved coverage rates and narrower predictive intervals  
  when we restricted to the upper bound of the prediction interval with failure times that were less than $\eta$ in the testing dataset.

We have also conducted a simulation study when the censoring times $C$ depend on covariate $\pmb X$.  
When the censoring distribution was correctly accounted for, the proposed method achieved the targeted 90\% coverage probability across all working models. Widths of the   intervals are slightly wider than their counter parts when $C$ is independent of $\pmb X$. 
It is somewhat surprising that even when the censoring times depend on $\pmb X$ but are ignored in the estimation procedure, the coverage rate could still maintain around 90\% when $T$ followed the log-normal distribution. There was a moderate undercoverage (around 85\%) when $T$ followed the Weibull distribution and the censoring distribution was mis-specified.

\begin{table}
\centering
  \caption{Empirical 90\% coverage  of the failure times when $C$ depends on $\pmb X$: when this dependence is ignored in the estimation using $G(t)$ and when it is properly considered using $G(t|\pmb X)$.}
    \begin{tabular}{lllccc} \toprule
    Distribution    &  Setting & Method & Cov  & Length & SD  
    \\  \midrule
    Weibull    & $G(t| \pmb X)$  &CPI-log-normal & 0.90  & 21.24 & 2.96 \\
               &  &CPI-Weibull & 0.90  & 21.18 & 2.85  \\
               &    &CPI-Cox &0.90  & 21.67 & 2.77  \\
         && &       &         \\
    Weibull &  $G(t)$ & CPI-log-normal & 0.84  & 16.72 & 1.32  \\
     &&CPI-Weibull &0.84  & 16.74 & 1.28  \\
     &&CPI-Cox &0.85  & 17.20 & 1.64 \\
        && &       &      \\
    Log-normal & $G(t|\pmb X)$ & CPI-log-normal& 0.90  & 70.64 & 19.59 \\
    && CPI-Weibull &0.90  & 70.79 & 20.48 \\
    && CPI-Cox &0.90  & 43.83 & 25.08 \\
      &&     &       &        \\
    Log-normal & $G(t)$ & CPI-log-normal& 0.90  & 68.07 & 15.39\\ 
    && CPI-Weibull &0.90  & 68.34 & 16.95 \\
    && CPI-Cox &0.91  & 41.41 & 16.20 \\
    \bottomrule
    \end{tabular}%
  \label{tab:rev1}%
\end{table}%

\section{Application to the breast cancer data from the NCDB database}

Using the proposed methods, we estimate the distribution-free conformal predictive   intervals for non-metastatic inflammatory breast cancer patient's survival outcome given a patient's tumour characteristics and initial treatment received. Patients who underwent surgical treatment during 1998-2010 were identified from the NCBD database. We ascertained 10,197 female patients who underwent surgical resection of any type with any histologic breast cancer subtype and any nodal status who had a clinical diagnosis of stage IIIB or IIIC IBC following the American Joint Committee on Cancer fifth or sixth editions. Among the 10,197 patients, 4,217 patients had complete information on death or censoring times, and covariates, age, race, Charlson-Deyo Score (CDS), tumour grade, nodal stage, and treatment. 
 Current IBC treatment guidelines recommend neoadjuvant chemotherapy followed by modified radical mastectomy and post-surgery radiation therapy, referred as trimodality treatment \citep{dawood2011differences}.   
 We randomly split patients into a training set (70\%) and a testing set (30\%) and calculated the coverage of the proposed method. As the actual failure time was only observed for the uncensored subjects, the algorithm for splitting conformal prediction to survival outcomes was employed to validate the overall coverage probability of CPIs before providing the estimated CPIs for patients with different risk factors.  We repeated the process of random splitting 100 times.
Applying the validation algorithm, it showed that the proposed methods (CPI-log-normal, CPI-log-Weibull, and CPI-Cox methods) achieved the specified coverage probability of 90\% (5th-95th percentile, 8.5-83.7 months), 90\% (8.3-82.8 months), and 90\% (8.3-79.5 months), respectively for the three CPIs. The estimated CPIs were comparable under three working models overall. In terms of coverage probability, the proposed CPIs methods are superior to other nonparametric or conditional approaches.  

\begin{figure}
\begin{subfigure}[b]{0.95\textwidth}
   \centerline{\includegraphics[width=1\linewidth]{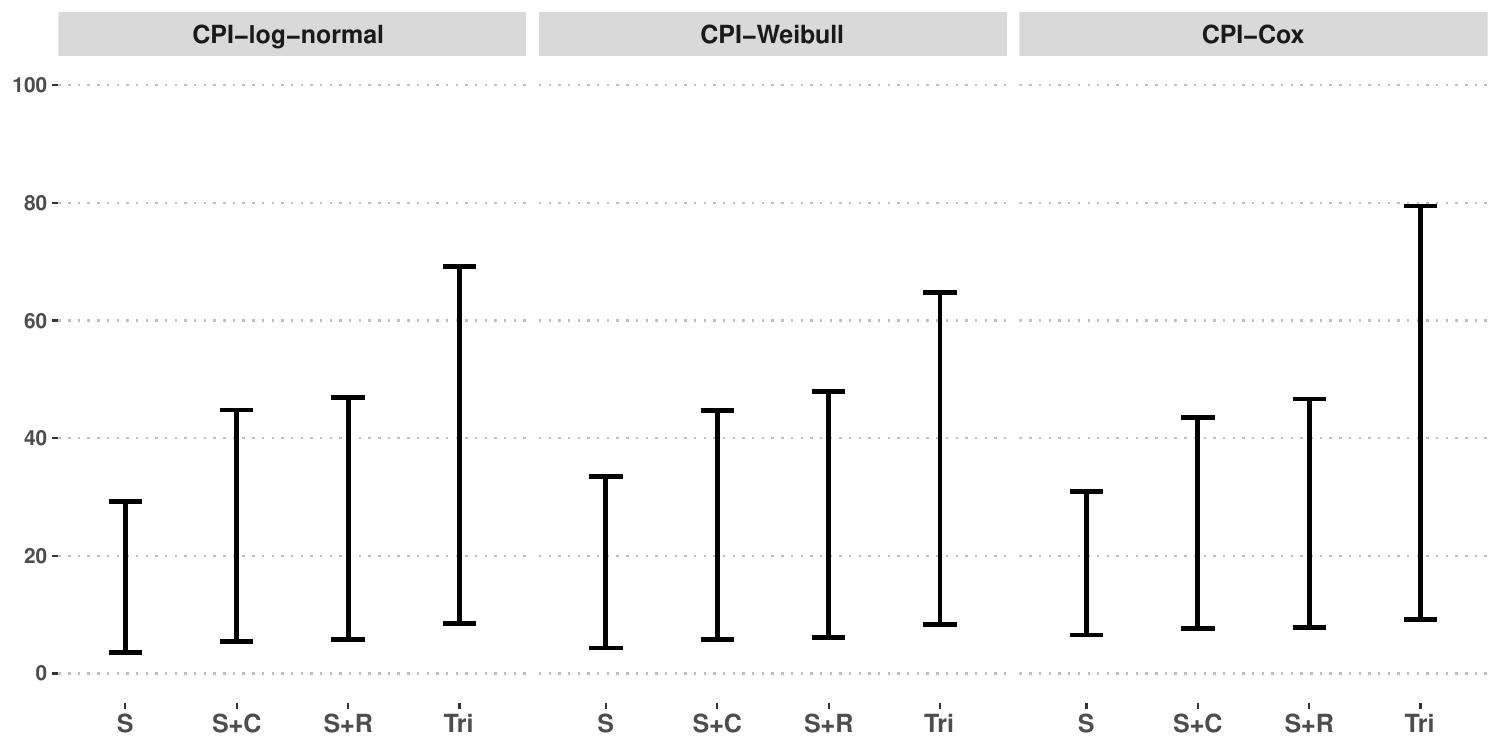}}
    \caption{Predicted 90\% coverage of the failure times in month for a 55-year-old white patient  without any co-morbidity condition, nodal stage N2, and poorly-differentiated tumour.}
    \label{fig4}
\end{subfigure}

\begin{subfigure}[b]{0.95\textwidth}
    \centerline{\includegraphics[width=1\linewidth]{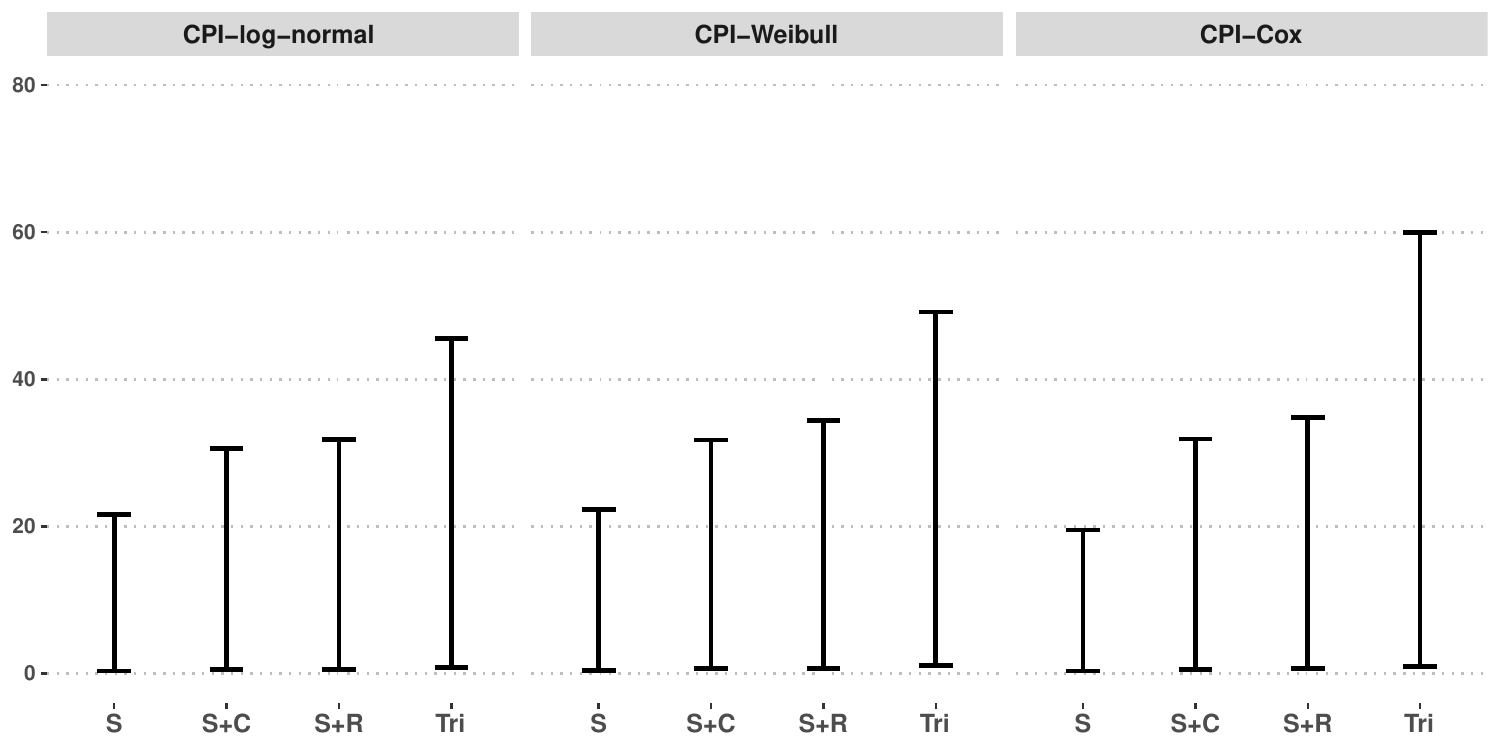}}
    \caption{Predicted 90\% coverage of the remaining failure times in month for a 55-
    year-old white patient  without any co-morbidity condition, nodal stage N2, and poorly-differentiated tumour who has survived 24 months after the initial IBC diagnosis.}
    \label{fig5}

\end{subfigure}
\caption{Predicted coverage of the failure times (a) or the remaining failure times (b) for a patient based on a given profile. The treatment options are: surgery alone (S), surgery plus chemotherapy (S+C), surgery plus radiation (S+R) and surgery, chemotherapy and radiation (Tri).}
\end{figure}

We then estimated the conformal predictive intervals for IBC patients and examined the impact of race, CDS, tumour grade, nodal stage, and especially treatment by varying the levels of variables. 
The conformal prediction intervals for patients with various profiles were estimated and shown in Figure \ref{fig4}. For example, it is of interest to estimate the prediction interval (with 90\% of coverage) of the survival time for a new patient with IBC diagnosis given her profile: 55-year-old white woman without any co-morbidity condition, nodal stage N2, and poorly-differentiated tumour. Figure \ref{fig4}  show that if she receives the trimodality treatment, her 90\% predictive survival time is in (8.5, 69.2), (8.3, 64.8), (9.1, 79.5) months under the log-normal, Weibull, or Cox working models, respectively. For the same patient, if she decides to receive surgery alone, her 90\% predictive survival time is in the range of (3.6, 29.3) or (4.3, 33.5) months under the log-normal or Weibull working models.  
Under the same treatment (e.g., trimodality), 
patients with lower CDS scores, well-differentiated or moderately differentiated tumours (compared to poorly- or undifferentiated tumours), and a lower nodal stage showed a higher low bound and a much higher upper bound, as shown in Table S5.
Though the estimated CPIs under the three working models are similar with less than one month difference at low bound, the upper bound for the CPI-Cox is much larger than the CPIs under the parametric working models. 

We recommend to use CPIs under the parametric working models given the heavy censoring rate (65\%) in this dataset. As shown in Figure 2, it is worth noting that the upper bound of the 90\% prediction intervals can be very different (around three to four years) under different treatment options, which carries critical information for patients and their clinicians to make more informed treatment plans.  On the other hand, the low bounds only differed one to four months given the different treatment plans.

 Given that a patient has survived for 24 months after her initial IBC diagnosis and treatment, it is of interest to predict her remaining survival time based on her tumour characteristics and treatment received as shown in Figure \ref{fig5}. For example,  a 55-year-old white patient with CDS (0), PD, and N2 who received trimodality treatment had a 90\% conformal predictive   interval of (0.8, 45.6) and (1.1, 49.2) months of remaining survival using the CPI-log-normal and CPI-Weibull approaches, respectively. This was compared to (0.3, 21.7) and (0.4, 22.3) months for the same subject if she had surgery alone, (0.5, 30.6) and (0.6, 31.8) months if she had surgery and chemotherapy, and (0.5, 31.8) and (0.7, 34.4) months if she had surgery and radiation.

\section{Concluding remarks}

While conformal inference has been successfully used in uncensored data and machine learning covariate shift problems, this method faces new challenges in predicting right-censored survival outcomes. 
One main difference between our proposed method and the method by \cite{candes2023conformalized} is the censoring mechanism.

In this paper, we propose a bootstrap method to mimic the pivot used for constructing the conformal predictive interval.
Conventionally, a pivotal quantity is a function of sample data and unknown parameters, but the distribution of the pivotal does not depend on the unknown parameters.  This property allows the pivotal quantity to be used for constructing  conformal intervals for the unknown parameters. In order to obtain a conformal predictive interval for the 
future survival outcome given covariates ${\pmb X =\pmb x}$,  we may perceive the conditional
survival probability $ S(T\mid {\pmb x}; {\pmb\theta})$ as the ``generalized pivot", which serves as a conformal score. Note that if $S(T\mid\pmb x;\pmb\theta)$ is correctly specified,  
$S(T\mid {\pmb x};{\pmb\theta})$ has a uniform distribution in $(0,1)$ conditioning on $\pmb X=\pmb x$. Subsequently, the unconditional distribution of $S(T\mid \pmb X; \pmb \theta)$ also follows the uniform $(0,1)$ distribution. A notable difference in conformal inference is that we consider the joint distribution of $(T,\pmb X)$ in the pivot, $S(T\mid \pmb X;\pmb\theta)$, instead of the conditional distribution in traditional inference. Estimating $P(S(T\mid \pmb X; \pmb\theta)\leq q)$ can be done using the empirical or weighted versions, while it would be difficult to estimate the conditional probability $P(S(T\mid\pmb X; \pmb\theta)\leq q| \pmb X=\pmb x)$ in traditional confidence interval approaches. 
Deriving the inverse estimate of the bivariate distribution by extending existing literature is relatively straightforward. However, a step of resampling paired data $(T_i, \pmb X_i)$, where $T_i$ is uncensored, from this consistently estimated distribution, represents a unique approach to constructing the conformal scores asymptotically. By using bootstrap replicates, the lower and upper quantiles for the conformal scores can be calibrated from the original data, taking into account the variation of both $T$ and $\pmb X$. Given that the pivot $S(t\mid{\pmb x};{\pmb\theta})$ is a monotone function of $t$, one can easily calculate the predictive interval for the future survival time $T_{n+1}$ given a set of covariates.

Recently, \cite{wang2021model} and \cite{zhang2023bootstrap} proposed to use the bootstrap method for constructing asymptotic prediction intervals; one essential difference with our method is 
their requirement that the conditional working model be correctly specified. 
\cite{cwiling2023comprehensive} constructed prediction intervals for a truncated version of the survival time. Different from our approach, they focused on the restricted mean survival time estimation and used a data splitting method. Since their  prediction intervals were derived from the weighted {mean residuals}, there was no guarantee that the lower bound would be positive.  We also conducted an additional simulation study to compare the performance of the method of \cite{cwiling2023comprehensive} with the proposed method. 
In summary, 
both methods were capable of achieving the desired 90\% coverage rate, and our proposed method led to smaller prediction interval widths. Moreover,  the method of \cite{chernozhukov2021distributional} could be generalized to survival outcome by splitting training data under required theoretical conditions and with sufficient large sample sizes. 

In our simulations, we have observed that achieving the nominal level of the lower one-sided predictive interval is relatively easier, regardless of the censoring rate or  underlying distributions (not presented here). However, attaining the nominal level for the right-side or a two-sided predictive interval 
is more difficult. This is especially true when the censoring proportion is high. This observation is not surprising, as it is well known that the K-M or Cox-model estimator cannot extrapolate beyond the last observation, i.e., at
the right tail when the censoring is high.

\section*{Conflict of interests}
No competing interest is declared for all authors.

\section*{Acknowledgments}
 The authors thank Ms. Jessica Swann from MD Anderson for her help in language editing.  

\section*{Funding}
This work is supported in part by funds from the National Institute of Health (R01CA269696 for J.N. and Y.S.) and U.S. National Clinical Trials Network Statistics and Data Center (U10CA180899 for P.J.).

\section*{Data Availability Statement}
The NCDB dataset used in the paper cannot be shared publicly due to our agreement with the NCDB for patients’ privacy protection. If readers are interested in obtaining access to the NCDB dataset for a specific study, please contact  NCDB directly at ncdb\_puf$@$facs.org,  to facilitate a data agreement with NCDB. 

\bibliographystyle{apalike}
\bibliography{reference.bib}

\end{document}